\documentclass[useAMS,usenatbib]{mn2e}

\usepackage{graphicx}

\title[Numerical simulations of downward convective overshooting in the giants]
      {Numerical simulations of downward convective overshooting in the giants}
\author[Chun-Lin Tian et al.]{Chun-Lin Tian$^{1,2}$\thanks{E-mail: cltian@bao.ac.cn},
Li-Cai Deng$^{1}$ and Kwing-Lam Chan$^{3}$\\
$^{1}$National Astronomical Observatories, Chinese Academy of Sciences,
      Beijing 100012, China\\
$^{2}$Graduate University of Chinese Academy of Sciences,
      Beijing 100049, China\\
$^{3}$Department of Mathematics, Hong Kong University of Science and
Technology,
      Hong Kong, China}
\begin{document}

\date{Accepted 2009 May 30. Received 2009 May 29; in original form 2008 July 15}

\pagerange{\pageref{firstpage}--\pageref{lastpage}} \pubyear{2009}

\maketitle

\label{firstpage}

\begin{abstract}
 An attempt at understanding the downward
overshooting in the convective envelopes of the post-main-sequence
stars has been made on the basis of three-dimensional large eddy
simulations, using artificially modified
 OPAL opacity and taking into account radiation and ionization
in the equation of state. Two types of stars, an intermediate mass
star and a massive star were considered. To avoid the long thermal
relaxation time of the intermediate mass star,
 we increased the stellar energy flux artificially
while trying to maintain a structure close to the one given by
one-dimensional stellar model.
 A parametric study of the flux factor was performed. For the
massive star, no such manner was necessary. Numerical results were
analyzed when the system reached the statistical steady state. It
was shown that the penetration distance in pressure scale heights is
of the order of unit. The scaling relations among the penetration
distance, input flux and vertical velocity fluctuations studied by
Singh et al. (1998) were checked. The anisotropy of the turbulent
convection and the diffusion models of third order moments
representing the non-local transports were also investigated. These
models are dramatically affected by the velocity fields and no
universal constant parameters seem to exist. The limitation of the
numerical results was discussed as well.
\end{abstract}


\section{Introduction}

Turbulent convection is one of the major uncertainties in our
understanding of stellar properties. As a very efficient mixing
process and energy transfer mechanism, it affects the stellar
structure and evolution significantly. Because of its high
non-linearity and complexity, we cannot solve the turbulent flows
analytically at present. Since \cite{emden07} established the first
practical model of stellar structure, a lot of effort has been made
to study the problems of stellar convection. A traditional treatment
is the mixing-length theory (MLT)(\cite{vitense58}). The basic idea
of MLT is that convection is viewed as rising and sinking bubbles
which blend into the surroundings after traveling a distance of $l$,
where $l=\tilde{\alpha} H_{\rm p}$, $\tilde{\alpha}$ is an arbitrary
parameter and $H_{\rm p}$ is the local pressure scale height. Since
the 1960's, several authors have worked directly with the
hydrodynamic equations (see \cite{spiegel71} and references
therein). By far, besides the various MLTs, the most practical
stellar convection theories are those based on the moment method
(\cite{xiong77,xiong86,xiong89a,canuto93}). The moment approach was
introduced first by \cite{keller24} and its first application in
stellar convection was discussed by Castor (1968, unpublished, see
the review by \cite{baker87}). In this method, the flow variables
are split into an average and a fluctuating part. Manipulation of
the hydrodynamic equations based on such splitting gives a set of
equations for the moments of the turbulent fluctuations. In these
equations, higher order moments appear and need to be approximated
by the so-called closure models. Most of these models use a gradient
type approximation, e.g. Xiong's theory. Along with the development
of digital computer and computational fluid dynamics (CFD),
numerical simulations have become an essential tool for the
astrophysical fluid dynamic community. The pioneer work in the
numerical study of stellar type convection is the two-dimensional
simulations of compressible convection done by \cite{graham75}.
Since the late 1980's, three-dimensional numerical experiments and
simulations have become the mainstream
(\cite{chan86,chan89,chan96,stein89,malagoli90,porter2000}).
Numerical investigation of turbulent flows has two categories:
direct numerical simulation (DNS) and large eddy simulation (LES).
The problem being three-dimensional, multi-length-scale and
multi-time-scale, a DNS of stellar convection demands enormous
computer resources. In LES, the large eddies are calculated
explicitly while the smaller eddies are to be handled by sub-grid
scale (SGS) models. This is usually the approach adopted for stellar
flow simulations. Being limited by the speed and memory of the
computers, the progresses made by numerical simulations are usually
slow. Even with today's machines, we still cannot simulate directly
the whole convection zone of a post-main-sequence star. A frequently
used approach is to make high resolution local simulation.

Many problems of stellar structure are associated with convective
overshooting. The overshooting distance beneath an outer convection
zone may affect the surface chemical composition, e.g., the
depletion of lithium in the Sun. The radiative-convective boundary
is thought to be closely related to the solar dynamo and is the site
for magnetic flux storage. The gravity waves generated by
penetration may generate momentum exchange with the interior.
However, calculation of overshooting is a challenge to all existing
stellar convection theories. Due to its intrinsic drawback, the
local MLT cannot handle the convective overshooting consistently. A
number of non-local models have been developed to incorporate the
feedback of the overshooting on the energy transport in the
penetration zone which is neglected in the local MLT. Although these
models can produce overshooting, \lq\lq they still consider
convection as an extended local phenomenon \rq\rq (see
\cite{roxburgh98} and references therein). In Xiong's hydrodynamic
stellar convection theory (1985b,1989b), the difficulties caused by
local treatment disappear spontaneously. Xiong's calculations show
extensive overshooting zones and the kinetic energy fluxes in such
zones are negligible. At the same time, the impact on the overall
stellar structure is minor (especially for zero-age-main-sequence
(ZAMS) stars). However, Xiong's closure scheme is under debate.
Numerical experiments can be used to test the assumptions used in
the one-dimensional models. \cite{roxburgh93} performed a
two-dimensional simulation of convective penetration to study the
integral constraints on the extent of overshooting by constructing a
temperature-dependent radiation conductivity model.
\cite{singh94,singh95,singh98} conducted a series of numerical
experiments to examine the scaling relationships among the
penetration distance ($\Delta_{\rm d}$), vertical velocity at the
bottom of the convective region and the total energy flux ($F_{\rm
b}$). The results of three-dimensional LESs of compressible
turbulent convection were used to test the relations proposed by
\cite{schmitt84} and \cite{zahn91}. They confirmed $\Delta_{\rm
d}\propto v_{\rm zo}^3/F_{\rm b}$ for nearly adiabatic penetration
and $\Delta_{\rm d}\propto F^{1/2}_{\rm b}$ for non-adiabatic
penetration where $v_{\rm zo}$ is the root mean square (rms)
vertical velocity at the unstable-stable interface. Using the same
technique, \cite{saikia2000} found that the numerical aspects of the
model, such as aspect ratio, grid number could greatly affect the
penetrative distance. \cite{brummell2002} performed a large number
of high-resolution, three-dimensional DNSs for the purpose of
investigating the penetration and overshooting of the turbulent
compressible convection. In their study, the effects of rotation
were included. Recently, penetration below a stellar type rotating
convection zone were estimated by \cite{pal2007} with a set of LESs.
 \cite{rogers2006} reported a two-dimensional simulation
of gravity waves induced by overshooting below the solar convection
zone. In these numerical experiments, only ideal gas with a
polytropic initial distribution was considered. However, for a
practical model, the realistic radiation opacity and the equation of
state(EOS) should be used.

According to the stellar evolution theory, at the base of the giants
branch in the Hertzsprung-Russell diagram, convection occurs in the
outer region of the envelope and extends to the deep stellar
interior. When the base of the convection zone overlaps with the
exhausted nuclear reaction area, the turbulent convection will
produce an efficient mixing and dredge up the processed material to
the surface of star. This mechanism is used to explain observed
surface chemical peculiarities at the surfaces of certain stars. In
this paper, we present a preliminary attempt to simulate
overshooting below the convective envelopes of some
post-main-sequence stars. Besides the EOS and the radiation opacity
(modified to accomodate an enhanced flux), the radiative pressure,
radiative energy were also taken into account. In Section 2, a brief
introduction to the adopted hydrodynamic code and input physics is
given. The construction of the initial hydrostatic models, the
simplifications and the parameters of the computational models are
also specified there. In Section 3 we analyze the numerical results
with a focus on the statistical properties of the turbulent
fluctuations, energy fluxes, overshooting, and moment closure. A
summary will be given in the Section 4 where the limitation and
extensibility of the current study are also discussed.

\section[]{Description of the Computational Model}

For simplicity, the magnetic field, rotation and nuclear
reaction are neglected. The motions of fully ionized gas in the
chemically homogeneous interior of stars are governed
by the Navier-Stokes (NS) equations:
\begin{eqnarray}
\label{ns1}
\partial\rho/\partial t &=& -\nabla \cdot \rho \vec{v},\\
\label{ns2}
\partial\rho \vec{v}/\partial t &=& -\nabla \cdot \rho \vec{v}\vec{v}
-\nabla p+\nabla\cdot\vec{\Sigma}+\rho\vec{g},\\
\label{ns3}
\partial E/\partial t &=&-\nabla\cdot[(E+p)\vec{v}-\vec{v}\cdot\vec{\Sigma}
+\vec{F}_{\rm d}]+\rho \vec{v}\cdot\vec{g},
\end{eqnarray}
where $\rho$ is the density, $\vec{v}$ is the velocity, $p$ is the
pressure, $\vec{g}$ is the gravitational acceleration, $E=e+\rho
v^2/2$ is the total energy where $e$ defined in Eq.(\ref{eos}) is
the internal energy  including ionization energy and radiation
energy.
\begin{equation}
\vec{\Sigma}=2\mu\vec{\sigma}+\lambda(\nabla\cdot\vec{v})\vec{I}
\end{equation}
is the viscous stress tensor, where $\vec{\sigma}$ is the strain
rate tensor, $\vec{I}$ is the identity tensor and $\mu$ represents
the sub-grid scale (SGS) eddy viscosity:
\begin{equation}
\label{mu} \mu=\rho(c_{\rm
\mu}\Delta)^2(2\vec{\sigma}:\vec{\sigma})^{1/2}.
\end{equation}
$c_{\rm \mu}$ is an adjustable constant parameter, $\Delta$ is
a length scale of the order of the local grid size, and the colon
represents the contraction of tensor. The bulk viscosity $\lambda$
is taken to be $-(2/3)\mu$.
\begin{equation}
\label{fd} \vec{F}_{\rm d}=\vec{F}_{\rm rad}+\vec{F}_{\rm sgs}
\end{equation}
is the diffusive flux. $\vec{F}_{\rm rad}$ is the radiative flux
which can be accurately computed as diffusion in the deep interior
of the star, i.e.,
\begin{equation}
\vec{F}_{\rm rad}=-\frac{4acT^3}{3\kappa\rho}\nabla T.
\end{equation}
$a$ is the radiation density constant, $c$ is the speed of light,
and $\kappa$ is the Rosseland mean opacity.
\begin{equation}
\label{fsgs} \vec{F}_{\rm sgs}=-\frac{\mu}{\sigma_{\rm sgs}}C_{\rm
p}(\nabla T-\nabla_{\rm a}\frac{T}{p}\nabla p)
\end{equation}
stands for the diffusive energy transport by the SGS turbulence.
$\sigma_{\rm sgs}$ is the effective Prandtl number for the SGS model
(hereafter, $\sigma_{\rm sgs}$ is called SGS Prandtl number),
$C_{\rm p}$ is the specific heat of the gas at constant pressure,
and $\nabla_{\rm a}$ is the adiabatic gradient.

The SGS Prandtl number $\sigma_{\rm sgs}$ is taken to be $1/3$; the
Deardorff number $c_{\rm \mu}$ in Eq.~(\ref{mu}) is fixed at $0.2$.
Near the solid boundaries, the SGS viscosity is enhanced to absorb
irrelevant acoustic waves.  In the stable region, the turbulent
diffusive flux (\ref{fsgs}) is taken to be zero. The equations
(\ref{ns1})$\sim$(\ref{ns3}) are solved in the Cartesian coordinates
$(x,y,z)$, where $z=(r-r_{\rm b})/d$ for $r \rm{(radius)}\in [r_{\rm
b},r_{\rm b}+d]$. $r_{\rm b}$ and $d$ are the radial location of the
bottom and the height of the computational domain, respectively (see
Section 2.3). The computed domain is a rectangular box with periodic
boundaries on the sides and solid boundaries at the top ($z=1$) and
bottom ($z=0$). The aspect ratio of the domain (width/height) is
$1.5$. The total energy flux $F_{\rm b}$ is fixed at the bottom; at
the top the entropy is fixed. The grid distribution is horizontally
uniform. In vertical direction, the grid spacing decreases smoothly
with the height. All cases use a $64\times 64\times 96$ mesh.

\subsection{Numerical Scheme: ADISM}

We adopted the ADISM (Alternating Direction Implicit method on
Staggered Mesh) scheme of \cite{chan82} to solve the full NS
equations (\ref{ns1})$\sim$(\ref{ns3}). This method has
second-order-accuracy in space which is adequate for simulating the
turbulent situation. As an implicit scheme, it can avoid the time
step restriction imposed by sound waves associated with the
Courant-Friedriches-Lewy(CFL) condition (see \cite{richtmyer1968}).
Numerical tests show that the ADISM scheme can maintain stability
for a time-step over 100 times the normal value suitable for
explicit schemes. In contrast with
 most other implicit methods, the CPU time consumption of the ADI
method is linearly proportional to the number of grid points. A
detailed examination of the ADISM approach has been given by
\cite{chan86}. The hydrodynamic code we used was initially developed
and used by \cite{chan89} and then adopted by \cite{kim98} to study
the upper solar transition layer. Therefore, it has been well
debugged.

\subsection{Input Physics: EOS and Opacity}

In the simulated region, the temperature is high enough that all
kinds of atoms are fully ionized. For a density $\rho$ and a gas
particle mass $Am_{\rm H}$, the Coulomb energy per particle is
$e_{\rm C}=\tilde{Z}^2\mathrm{e}^2/(4\pi\epsilon_0\mathcal{D})$,
where $\tilde{Z}\mathrm{e}$ is the particle charge, $\epsilon_0$ is
the permittivity, $\mathcal{D}=(Am_{\rm H}/\rho)^{1/3}$ is the mean
inter-particle distance, $m_{\rm H}$ is the atomic mass unit.
 At temperature $T$, the kinetic energy per particle has the form $e_{\rm k}=3kT/2$,
 where $k$ is the Boltzmann constant.
The ratio of Coulomb energy to kinetic energy can be estimated as
\begin{equation}
\frac{e_{\rm C}}{e_{\rm
k}}=\left[\frac{\tilde{Z}^2\mathrm{e}^2}{6\pi\epsilon_0 k(Am_{\rm
H})^{1/3}}\right] \frac{\rho^{1/3}}{T},
\end{equation}
where $A$ and $\tilde{Z}$ are of the order of unit, the term in
square brackets is approximately $10^5$. For our modeled cases, the
ratio of $\rho^{1/3}/T$ is less than $10^{-6.5}$, which means
$e_{\rm C}/e_{\rm k}\approx 0.01$. This indicates that the particle
interactions are dominated by collisions and the nonideal effects
caused by the Coulomb force can be safely neglected. Hence, for the
present calculation, the gas is treated as a fully ionized ideal
gas. The internal energy $e$ and total pressure $p$ can be expressed
as:
\begin{equation}
\label{eos}
e=aT^4+\frac{3}{2}p_{\rm g}+\rho e_{\rm i},\\
p=p_{\rm g}+p_{\rm r},
\end{equation}
respectively, where $p_{\rm g}=\beta p$ is the gas pressure, $p_{\rm
r}$ is the radiation pressure, and
 $e_{\rm i}$ is the ionization energy per mass unit.
Using Eqs. (\ref{eos}) we can conveniently calculate thermodynamic
variables such as $C_{\rm p}$, $\nabla_{\rm a}$, $(\partial
T/\partial \rho)_e$, etc. Comparisons show that the discrepancies
between the $\rho$, $C_{\rm p}$ and $\nabla_{\rm a}$ calculated by
equation (\ref{eos}) and the interpolation from the OPAL EOS tables
(\cite{rogers96}) are less than $2\%$ in the regions we computed.

The opacity $\kappa$ is obtained by interpolating the OPAL tables
(\cite{rogers92}). Before interpolation, the OPAL opacity is
tabulated as functions of density and $\mathcal{R}=\rho/(T\times
10^{-6})^3$ in two-dimensional arrays with even $\Delta \ln{\rho}$
and $\Delta \ln{\mathcal{R}}$. A \emph{hunt} method is used to
search the table.

\subsection{Initial Models: Construction and Modifications}

\begin{table}
 \caption{Key properties of the reference stars.}
 \label{key}
 \begin{tabular}{@{}ccccc}
  \hline
  $M/M_\odot$ & $\log{(T_e)}$ & $\log{(L/L_\odot)}$
        & Age($yrs$)
        & $R/R_\odot$ \\
  \hline
  3  &3.695 & 2.032 & $4.258\times 10^8$ & 14.113\\
  15 &3.761 & 4.869 & $1.308\times 10^7$ & 273.376\\
  \hline
 \end{tabular}

 \medskip
 Both of these two stars have the same chemical composition
: $X=0.7, Z=0.02$.
\end{table}

We consider two types of post-main-sequence stars: a massive star of
15$M_\odot$ (red super giant) and an intermediate mass star of
3$M_\odot$ (red giant). The regions we study are far away from the
core and surface, so that we need not consider the Non-LTE (local
thermal equilibrium) effects and nuclear reaction. Our 1D reference
stellar models are computed with more realistic input physics and
Xiong's non-local time dependent
 stellar turbulent convection theory (private communication). The model
properties (i.e., mass $M$, effective temperature $T_e$, luminosity
$L$, age and radius $R$) are listed in Table~\ref{key}.

In the simulation of the intermediate mass star envelope, we have
encountered serious difficulties. They are discussed as follows.
\begin{enumerate}
 \item The first problem is the very long duration of thermal
relaxation. The thermal relaxation time scale can be estimated by
dividing the total energy contained in the system by the rate of
energy input at the bottom. For intermediate mass star, the
dimensionless input energy flux ($F_{\rm b}$) is of the order of
$10^{-5}$ and the radial integral of energy density energy ($e$)
above the bottom ($z=0$) is on the order of $1000$. This means that
the system needs a time scale of $10^8$ to reach a thermally relaxed
state. The time step achieved by the ADISM scheme is around
$10^{-2}$ for current study. Therefore, we need about $10^{10}$
steps to accomplish the relaxation. This requires too much
computational resources.
\item Meanwhile, the limited spatial resolution is unable to
handle the small superadiabatic gradient
($\Delta\nabla={\partial\ln{T}}/{\partial\ln{p}}-\nabla_{\rm a}$)
accurately. In the efficient convective region, $\Delta\nabla$ is on
the order of $10^{-6}$, which would be easily overwhelmed by the
numerical truncation errors.
\end{enumerate}

A way to go around the above problems is to enlarge the total energy
flux (and thus the superadiabatic gradient) to make the computation
feasible. The stellar thermal structure and convective stability are
mainly determined by the energy transport mechanism and the
superadiabatic gradient. Using the expression of the diffusive
radiative flux, we have
\begin{eqnarray}
\label{cstr1}
\frac{d T}{d z}&=&-\frac{3\kappa \rho F_{\rm rad}}{4acT^3},\\
\label{cstr2} \frac{d p}{d z}&=&\frac{p}{(\Delta\nabla+\nabla_{\rm
a})T}\frac{d T}{d z}=\rho \vec{g},
\end{eqnarray}
where
\begin{eqnarray}
\nabla_{\rm a}&=&\nabla_{\rm a}(p,T),\\
\kappa&=&\kappa(p,T),\\
\rho&=&\rho(p,T).
\end{eqnarray}
Given the distributions of $F_{\rm rad}$ and $\Delta\nabla$, the EOS
and opacity, and the values of $p$, $T$ at one boundary,
 we can integrate (\ref{cstr1}, \ref{cstr2}) to construct a stratified layer of
fluid close to the reference stellar model.

In order to reduce the thermal relaxation time for the intermediate
mass star model, the radiative flux $F_{\rm rad}$ (=$F_{\rm b}$ in
the radiative region) must be enlarged dramatically (by a factor of
$a_{\rm f}\gg 1$). This would make the thermal structure calculated
by equations (\ref{cstr1}, \ref{cstr2}) totally different from the
reference stellar structure. This problem can be solved by dividing
the opacity the same factor $a_{\rm f}$. In doing so, $P$, $T$ and
the profile of $F_{\rm rad}/F_{\rm b}$ are similar to the reference
stellar model and only the magnitude of $F_{\rm rad}$ is enhanced.
In the interior of stars, for constant luminosity, $F_{\rm rad}$ is
inversely proportional to the square of radius $r$. If the
structures calculated in spherical coordinates are into the
Cartesian frame, the systems will undergo large adjustments and will
substantially deviate from the initial 1D stellar models. The
radiation conductivity is multiplied by a factor $c_{\rm r}=(r_{\rm
b}+z)^2/r_{\rm b}^2$ to avoid such adjustment. Here $r_{\rm b}$ is
the distance from the bottom of computed domain to the stellar
center. Hence, for the calculations, we use an effective opacity
$\kappa^\ast=\kappa/(c_{\rm r} a_{\rm f})$ instead of the real
$\kappa$.

\begin{figure}
  \includegraphics[width=8.0cm]{./figures/ini-flx.eps}\\
  \includegraphics[width=8.0cm]{./figures/ini-sgt.eps}
  \caption{Initial distributions of radiative flux and
  superadiabatic gradient
  for intermediate mass star (IMS) model (solid lines)
  and massive star (MS) model (dotted lines).
  The MS model is taken from 1D stellar model and
  the IMS model is constructed according to the 1D stellar model.
  Case I1, I2 and I3 listed in Table~{\ref{runs}} has the same initial $\Delta\nabla$ and
  different initial $F_{\rm rad}$.
  The solid line  in upper panel is an example of initial distribution of
  $F_{\rm rad}$ for IMS model ($2\times F_{\rm rad}$ of Case I3).
  The dashed line is the fake stable layer
  introduced to prevent the dramatic effects near the upper boundary.
  The abscissa is the height from bottom $z=(r-r_{\rm b})/d$ for $r\in [r_{\rm b}, r_{\rm b}+d]$,
  where $d$ is height of the computed zone
  and $r_{\rm b}$ is the location of the bottom.}
  \label{init}
\end{figure}

\begin{table*}
 \centering
 \begin{minipage}{150mm}
  \caption{Definition of physical model.}
  \label{dpm}
  \begin{tabular}{@{}lccccccccr@{}}
  \hline
 Type of star& $\log{(p_{\rm scl})}$ & $\log{(T_{\rm scl})}$ & $\log{(\rho_{\rm scl})}$ &
 $p_{\rm btm}$ & $T_{\rm btm}$ & $\rho_{\rm btm}$ &
 $d/R$ &$r_{\rm b}/d$&$F_{\rm btm}$\\
 \hline
 Intermediate& 7.881 & 4.872 & -5.116 & 737.152& 6.226
  & 118.276 & 0.1&6 & $0.2\times 10^{-4}$ \\
 Massive     & 5.820 & 5.022 & -7.602 & 927.728& 5.095
  & 226.088 & 0.379&1 & 0.651 \\
\hline
\end{tabular}

\medskip
 $p_{\rm btm}$, $T_{\rm btm}$, $\rho_{\rm btm}$ are scaled by
 $p_{\rm scl}$, $T_{\rm scl}$, $\rho_{\rm scl}$ respectively.
 The input flux $F_{\rm btm}=F_{\rm b}/(c_{\rm r}a_{\rm f})$ is scaled by
 $(p_{\rm scl}^{1.5}/d_{\rm scl}^{0.5})$, where $F_{\rm b}$ is the modified dimensionless input
 flux.

\end{minipage}
\end{table*}

An initial stratified layer constructed by the above method is
specified by the six parameters: $\rho_{\rm t}$, $p_{\rm t}$,
$T_{\rm t}$, $a_{\rm f}$, $r_{\rm b}$ and $d$ (the subscript t
denotes values at the top), and two distributions: $F_{\rm rad}$ and
$\Delta\nabla$. Their values are chosen to make the initial state
close to the reference model.

In the interior of the massive star, convection is not very
efficient. The zone extends for about $3.41$ pressure scale heights
(PSHs) ($0.25$ of the stellar radius $R$). The upper convective
boundary is located in the deep interior. The typical value of
$\Delta\nabla$ is on the order of $0.001$ in the convection zone,
while about $-0.06$ in the stable region below. The computed domain
of our massive star model contains about $6.8$ PSHs ($d=0.379R$)
total. It includes the whole convection zone and a lower stable
layer. The upper boundary is a little higher than the actual top of
the convection zone of the 1D reference model. The distance between
the bottom of computed domain and the center of the star is $r_{\rm
b}=0.379R$. The initial distributions of the radiative flux $F_{\rm
rad}$ and superadiabatic gradient $\Delta\nabla$ for the various
models are shown in Fig.~\ref{init} by dotted lines without
amplification factors (e.g., $a_{\rm f}=1$). The other
characteristics of the initial models are given in Table~\ref{dpm}.

For the massive star, the values of $F_{\rm btm}$ and other
quantities are computationally manageable and $a_{\rm f}$ can be
chosen to be $1$. On the other hand, the parameters of the
convection zone of the intermediate mass star have to be modified to
make the calculations feasible. We use several values of $a_{\rm f}$
to check for the parametric behaviors of the modified models.
 The convection in the red
giant extents to a height of about $14$ PSHs ($0.4R$). A direct
simulation of such a deep layer is unaffordable even with an
enlarged energy flux. We can only consider the lower part of the
convection zone. The computed box has a shorter radial extension
($d=0.1R$) and the top is placed at the middle of the convection
zone. $r_{\rm b}$ is at $0.6R$. Consequently, the unstable layer is
about $2.7$ PSHs thick and the whole computed domain contains about
$6.6$ PSHs.
 To suppress the
significant thermal adjustment caused by truncating the convection
zone, an artificial thermal control layer is introduced near the
upper boundary. This is a slightly sub-adiabatic layer with an
artificial conductivity set to $-F_{\rm rad}/(d T/d z)$ so that
radiation carries all of the energy flux. The corresponding layer is
indicated by the dashed line in Fig.~{\ref{init}}.

The dimensionless input energy flux ($F_{\rm btm}$) for intermediate
mass star model is very small (see the last column in
Table~\ref{dpm}). To reduce the thermal relaxation time several
values of $a_{\rm f}$ around $10^4$ are used. In constructing the
initial model, the flux and geometry were modified through the
controlling parameters $a_{\rm f}$, $c_{\rm r}$, $d$, and $r_{\rm
b}$. Among the modifications, we care most about the effects caused
by $a_{\rm f}$.
 Three intermediate mass star models are constructed
with different values of $a_{\rm f}$ (see Table~\ref{runs}).
 If $a_{\rm f}$ is much greater than that
of Case I3, the program can easily
 crash, and a value smaller than that of Case I1 would make the
 results affected by the numerical errors
(truncation and round-off errors). In Table~{\ref{runs}} $F_{\rm b}$
is the modified dimensionless numerical input energy flux, i.e.,
$F_{\rm b}=F_{\rm btm}c_{\rm r}a_{\rm f}$. The initial
superadiabatic gradient and radiation flux for the intermediate mass
star models are shown in Fig.~\ref{init}. The three models have the
same initial $\Delta\nabla$ (see the solid line in bottom panel of
Fig.~\ref{init}). Their $F_{\rm rad}$ have the same shape but
different magnitudes, the top panel of Fig.~\ref{init} just gives an
example of radiation flux distribution. Two times the initial
$F_{\rm rad}$ of Case I3 is plotted for clarity. Note that the
constructed intermediate mass star models are based on local
convection theory, i.e., there is no overshoot hump in the radiative
flux distribution (compare with Fig.5 of \cite{xiong01}, hereafter
XD2001).

\begin{table}
 \begin{center}
 \caption{Characters of the numerical runs.}
 \label{runs}
 \begin{tabular}{@{}lclcr}
  \hline
  Identifier & $a_{\rm f}$ &$F_{\rm b}$ & $t$
        & $\sigma_{\rm max}$ \\
  \hline
  I1  &$1\times 10^4$ & 0.05 & 9295 & 5.9\%\\
  I2  &$2\times 10^4$ & 0.10 & 7316 & 1.7\% \\
  I3  &$4\times 10^4$ & 0.20 & 5844 & 1.3\% \\
  M1  &1.0            & 0.651 & 11298 & 1.8\% \\
  \hline
 \end{tabular}
\end{center}

$a_{\rm f}$, $F_{\rm b}$, $t$ and $\sigma_{\rm max}$ are the
amplifying parameter, dimensionless input energy flux, dimensionless
relaxation time and maximum deviation of average total energy flux
from input flux, respectively.

\end{table}

For convenience, we make all the quantities dimensionless by scaling
the variables so that $\rho_{\rm t}$, $p_{\rm t}$, $T_{\rm t}$ and
$d$ all have the value one. The scalings, $\rho_{\rm scl}$, $p_{\rm
scl}$ and $T_{\rm scl}$ are given in Table~\ref{dpm}. $d/R$
represents the radial fraction of domain included in the
computation, and the last column gives the dimensionless energy
fluxes of the unmodified models. All the scalings of the other
variables can deduced from dimensional analysis.

\section{Results and Discussions}

\begin{figure}
  \includegraphics[width=8.2cm]{./figures/mm10-flxs.eps}
  \includegraphics[width=8.2cm]{./figures/m15-flxs.eps}
  \caption{Height distributions of average fluxes for Case I2 (top)
   and Case M1 (bottom). The enthalpy flux,
  kinetic flux, viscous flux, diffusive flux
  and total flux are represented by dotted lines,
  dashed lines, dash dot lines, dash dot dot lines
  and solid lines, respectively. Double-headed arrow indicates the overshooting zone, defined by negative
  convective flux.
  The vertical dashed lines represent the integral
  pressure scale heights counted from the upper boundary.}
  \label{flx}
\end{figure}
\begin{figure}
  \includegraphics[width=8.2cm]{./figures/sags-im.eps}
  \includegraphics[width=8.2cm]{./figures/sag-ms.eps}
  \caption{Height distribution of superadiabatic gradient
   for Case I1 -- I3 (top) and Case M1 (bottom). In the
top panel, solid line: I1; dotted line: I2; dashed line: I3.
Double-headed arrow indicates the overshooting zone. The vertical
dashed lines represent the integral
  pressure scale heights counted from the upper boundary.
  The very large superadiabatic gradients
  near the upper boundary are mainly caused by the solid boundary conditions.
  In the middle region of convection zone, the small superadiabatic
  gradient could be affected by the numerical errors. (see text for details)}
  \label{sags}
\end{figure}

After a long period of thermal relaxation, the flow reaches a
statistically steady state and its average properties become
stationary. One criterion of thermal relaxation is that the input
energy flux from the bottom is balanced by  the outgoing energy flux
through the top surface. In the present calculations, the deviations
of the averaged vertical total flux from the input energy flux are
less than $2$\% everywhere (except Case I1 which has a bound of
$6$\%).
 The maximum discrepancies
occur near the unstable-stable interfaces (where the
cross-correlation of temperature and vertical velocity changes its
sign).

In expressing the results, the overline denotes a combination of
horizontal and temporal averaging, the prime ($'$) denotes the
deviation from the mean, and the double prime ($''$) denotes the rms
fluctuation from the mean. In some of the figures, the locations of
integral pressure scale heights counted from the upper boundary are
also shown. For example, in Fig.~\ref{flx}, the first vertical
dashed line from right is $1$ PSHs away from the top boundary.
$H_{\rm p}=-d r/d\ln{p}$ sometimes does not increase monotonously
with depth
 as the gravity $g$ is variable and the EOS has a complex form.
In some plots (e.g., Fig.~\ref{flx}), the overshooting zone
$\Delta_c$, defined in Section \ref{sec:ovsh}, is indicated by a
double-headed arrow.

\subsection{Stationary Solutions}

\begin{figure}
  \centering
  \includegraphics[width=8.2cm]{./figures/vfyz-im.eps}
  \includegraphics[width=8.2cm]{./figures/vfyz-ms.eps}
  \caption{Velocity field projected onto the $x$--$z$ plane at $y=0.75$.
   Upper panel: Case I2; lower panel: Case M1.
   The regions confined by solid lines indicate the overshooting zones,
   i.e., $\Delta_c$ defined in Section \ref{sec:ovsh}.}
  \label{vfs}
\end{figure}

Figure~\ref{flx} shows the distributions of averaged vertical fluxes
for Case I2 and Case M1, from which we can see that the systems are
almost completely relaxed. The different energy fluxes are
defined as follows:
\begin{equation}
F_{\rm e}=\overline{v_{z}(e+p)}
\end{equation}
is the enthalpy flux,
\begin{equation}
F_{\rm k}=\frac{1}{2}\overline{v_{z}\rho v^2}
\end{equation}
is the flux of kinetic energy,
\begin{equation}
F_{\rm v}=\overline{v_{i}\Sigma_{iz}}
\end{equation}
is the small-scale eddy viscosity flux, where the Einstein summation
convention is used. The diffusive flux $F_{\rm d}$ is defined in
(\ref{fd}) and the total flux is $F_{\rm t}=F_{\rm e}+F_{\rm
k}+F_{\rm v}+F_{\rm d}$. A more detailed discussion of these fluxes
is given in Section~\ref{sec:flx}.

As an important indicator of convective instability, the
superadiabatic gradient $\Delta\nabla$ is given in Fig.~\ref{sags}
for all the cases. Here we focus on cases I1--I3. Highly
superadiabatic regions
 occur near the upper
stable-unstable interfaces, the systems are substantially unstable.
The sharp jumps may be caused by the lack of enough room for upward
motions. The same phenomenon was found in \cite{singh98}'s work (see
Fig.5 therein). In the convection zone, $\Delta\nabla$ is not as
small as that of the 1D reference model. It is caused by the
enhanced total flux. The slightly negative $\Delta\nabla$ in the
lower part of the convection zone ($0.4 < z < 0.6$) indicates
$\Delta\nabla$ is sensitive to even small numerical inaccuracy. A
higher-order scheme with denser mesh may eliminate such problem.

Figure~\ref{vfs} shows examples of the velocity fields projected
onto a vertical plane. We can see from these figures that the
turbulent flows are dominated by large eddies whose sizes are
comparable to at least a press scale height. The small scale flows
are mostly associated with downward plumes. The penetrations of such
plumes in to the overshoot zone are also shown. The upper solid
lines in Fig.~\ref{vfs} (near $z=0.4$ for Case I2, and $z=0.3$ for
Case M1) show the convective boundary where the enthalpy fluxes
change sign. The regions between two horizontal solid lines are the
overshooting zones with depths $\Delta_{\rm c}$ defined in Section
\ref{sec:ovsh}. In the lower stable layer, the motions are mainly
horizontal. This is characteristic of flows in convectively stable
regions.

\subsection{Fluctuations}
\label{sec:fluc}
\begin{figure}
  \centering
  \includegraphics[width=8.2cm]{./figures/dpt-f10.eps}
  \includegraphics[width=8.2cm]{./figures/dpt-ms.eps}
  \caption{Height distributions of relative fluctuations
 of density, pressure and temperature
 for Case I2 (upper panel) and Case M1 (lower panel).
Solid lines: $\rho''/\overline{\rho}$; dotted lines:
$p''/\overline{p}$; dashed lines: $T''/\overline{T}$. Double-headed
arrow indicates the overshooting zone. The vertical dashed lines
represent the integral
  pressure scale heights counted from the upper boundary.}
  \label{dpt}
\end{figure}
\begin{figure}
  \centering
  \includegraphics[width=8.2cm]{./figures/drms-scl.eps}
  \includegraphics[width=8.2cm]{./figures/prms-scl.eps}
  \includegraphics[width=8.2cm]{./figures/trms-scl.eps}
  \caption{Comparison of scaled fluctuations of thermodynamic variables,
   i.e., $\rho''/(\overline{\rho}F_{\rm b}^{0.5}$),
   $p''/(\overline{p}F_{\rm b}^{0.5}$) and
   $T''/(\overline{T}F_{\rm b}^{0.5}$). Solid lines: Case I1; dotted lines:
Case I2; dashed lines: Case I3.}
  \label{dpt-scl}
\end{figure}
\begin{figure}
  \includegraphics[width=8.2cm]{./figures/vxrms-scl.eps}
  \includegraphics[width=8.2cm]{./figures/vzrms-scl.eps}
  \caption{Comparison of scaled fluctuations of
  $v''_x$($v''_x/F_{\rm b}^{0.25}$)
  and $v''_z$ ($v''_z/F_b^{0.25}$).
  Solid lines: Case I1; dotted lines: Case I2; dashed lines: Case I3.}
  \label{vrm-scl}
\end{figure}

Two examples (Case I2 and Case M1) of the relative fluctuations of
the thermodynamical variables are shown in Fig.~\ref{dpt}, where
$\rho''/\overline{\rho}$, $p''/\overline{p}$ and $T''/\overline{T}$
are represented by the solid lines, dotted lines and dashed lines,
respectively. In the intermediate mass star models, the relative
fluctuation of temperature is comparable to that of density. In the
overshooting region, the temperature fluctuation shows a hump. In
massive star model where convection is inefficient,
 $T''/\overline{T}$ is substantially smaller than
$\rho''/\overline{\rho}$. Compared to $\rho''/\overline{\rho}$ and
$T''/\overline{T}$, the distribution $p''/\overline{p}$ is
relatively smooth.

The scaling relationships between the relative fluctuations and the
input energy fluxes are of particular interest. Figure~\ref{dpt-scl}
illustrates them for our models. The relative fluctuations of
thermodynamical variables are scaled by $F_{\rm b}^{0.5}$.
Figure~\ref{vrm-scl} illustrates the scaling relations between the
velocity fluctuations and the input flux. The velocity fluctuations
are scaled by $F_{\rm b}^{0.25}$.
 Away from the lower overshooting region,
these kinds of scaling relationships are good.
 Only Case I3 shows some small shifts in the curves. In this case
 turbulent convection is not efficient as those in
Case I1 and Case I2. In the overshooting region, the vertical
velocity fluctuations prefer the scaling $v''_z\propto F_{\rm
b}^{1/3}$. The reason why the current scaling relationships (eg.,
$v''_z\propto F_{\rm b}^{0.25}$) are different from those of
\cite{chan89}(hereafter CS89) (eg., $v''_z\propto F_{\rm b}^{1/3}$)
 is preliminarily the influence of radiation. Convection in the
 current models are not efficient as those in CS89. Radiation
 diffusion plays a very significant role even in the convection
 zone.

\subsection{Fluxes}
\label{sec:flx}

\begin{figure}
  \centering
  \includegraphics[width=8.2cm]{./figures/e-flxs.eps}
  \includegraphics[width=8.2cm]{./figures/k-flxs.eps}
  \includegraphics[width=8.2cm]{./figures/d-flxs.eps}
  \caption{Comparison of scaled fluxes for
  intermediate mass star model. (a)$F_{\rm e}/F_{\rm b}$;
  (b)$F_{\rm k}/F_{\rm b}^{0.75}$; (c)$F_{\rm d}/F_{\rm b}$.
Solid lines: Case I1; dotted lines: Case I2; dashed lines: Case I3.}
  \label{flx-scl}
\end{figure}
As illustrated in Fig.~\ref{flx}, different mechanisms dominate the
energy transport in different regions of the computed models. In
 the stable regions, the energy is predominantly transported by radiation.
The enthalpy flux and kinetic flux are the dominant modes of energy
transport in the convection zones. In the overshooting layer energy
transport is amidst the counterbalance of convection and radiation.
The hump in the distribution of radiative flux is to balance the
negative enthalpy flux in the overshooting layer. The relative
amplitudes of these humps are about 5 times those based Xiong's
theory (see Fig.5 in XD2001). This may simply reflect the fact that
total energy flux has been artificially enhanced.

Figure~\ref{flx-scl} compares the different types of energy fluxes
scaled by the powers of input flux. Unlike the results of CS89,
$F_{\rm e}$ and $F_{\rm k}$ are not scaled by $F_{\rm b}$ here. The
kinetic fluxes are approximately scaled by $F_{\rm b}^{0.75}$, which
could be regarded as a consequence of ${v''_z}\propto F_{\rm
b}^{0.25}$ since $F_{\rm k}\propto{v''_z}{v''}^2\propto {v''_z}^3$,
where ${v''}^2={v''_x}^2+{v''_y}^2+{v''_z}^2$. Figure~\ref{flx-scl}
(c) shows that in the convection zones, $F_{\rm d}$ is not scaled by
$F_{\rm b}$. This is also a consequence of significant effects of
radiation. High non-linearity makes the analysis of such phenomenon
very difficult.

Since $F_{\rm rad}$ is amplified by a factor of $a_{\rm f}\sim
O(10^4)$ the flux parameter is far from the correct value. Although
we have constructed three cases with different $a_{\rm f}$ to check
for the scalings, the results are restricted by the limited range of
values accessible by numerical simulations. The extent of the
computed domain and the boundary conditions can also affect the
relationships. All these make the applicability of the scaling
relations debatable. A comprehensive investigation of these
uncertainties needs a large number of models and massive
computations. It is beyond the scope of the present study.

\subsection{Overshooting}
\label{sec:ovsh}
 Ideally, the overshooting $\Delta_{\rm d}$ is
measured by the distance from instability boundary
($\Delta\nabla=0$) to the place where the transport velocity
vanishes ($v_{z}=0$). In reality due to the various numerical
limitations, this is impractical. Following \cite{singh95}, we
define $\Delta_{\rm d}$ to be the location at which $F_{\rm k}$
falls to $5\%$ of its value at the stable-unstable interface. An
alternative choice has been suggested by \cite{deng2007}. They
argued that the overshooting distance should be defined as the depth
of the region where the convective enthalpy flux is negative. We
also consider this length and denote it by $\Delta_{\rm c}$.
Numerical estimates of the overshooting distances are listed in
Table~\ref{ovsh}.

As we mentioned earlier, in the thermally relaxed systems, there is
no extensive adiabatic layer. For each of the  cases, a thin layer
with highly superadiabatic stratification exists near the upper
unstable-stable interface. $\Delta\nabla$ changes its sign in the
middle of the computed domain but remains close to zero till the
lower unstable-stable interface of the initial model is reached.
$\Delta\nabla$ then becomes substantially negative (see
Fig.~\ref{sags}). Therefore, we measure $\Delta_{\rm d}$ as the
distance between the \emph{initial} unstable-stable interface and
the location where $F_{\rm k}$ has fallen to 5\% its value at the
initial stable-unstable interface. Similarly, the vertical velocity
fluctuation at the convective boundary $v''_{\rm zo}$ is calculated
at the initial unstable-stable interface. The upper boundary of
$\Delta_{\rm c}$ is the convective boundary where enthalpy flux
changes its sign. For Case M1 the lower boundary of $\Delta_{\rm c}$
is at where $F_{\rm e}$ changes its sign again. For Case I1--I3, it
is at the layer where $|F_{\rm e}|=5\% |\min{(F_{\rm e})}|$. Since
$\Delta_{\rm d}$ is substantially less than $\Delta_{\rm c}$ and is
affected by the vertical resolution, we study the scaling
relationships among $\Delta_{\rm c}$, ${v''}_{\rm zo}$ and $F_{\rm
b}$ instead of $\Delta_{\rm d}$, ${v''}_{\rm zo}$ and $F_{\rm b}$.

\begin{table}
 \centering
 \caption{Overshooting distance and its scaling relationships.}
 \label{ovsh}
 \begin{tabular}{@{}lcccr}
  \hline
  Model & $\Delta_{\rm d} $ (in ${\rm PSHs}$) & $\Delta_{\rm c}$ (in ${\rm PSHs}$)
        & $\Delta_{\rm c}/{v''}_{\rm zo}^{3/2}$ & $\Delta_{\rm c}/F_{\rm b}^{1/2}$\\
  \hline
  I1  & 0.120 (0.910) & 0.140 (1.117) &  2.731 & 0.626 \\
  I2  & 0.124 (0.956) & 0.194 (1.658) &  2.952 & 0.614 \\
  I3  & 0.128 (0.993) & 0.241 (2.058) &  3.065 & 0.541 \\
  M1  & 0.224 (1.674) & 0.139 (0.856) &  2.374 & 0.173 \\
  \hline
 \end{tabular}

 \medskip

 PSHs stands for pressure scale heights.
 ${v''}_{\rm zo}$ is the vertical velocity
 fluctuation at the initial stable-unstable interface.
 All of them are dimensionless values.

\end{table}

The overshoot extents from present study are comparable to the local
 pressure scale heights and well scaled by ${v''}_{\rm zo}^{3/2}$.
The scaling relationship between $\Delta_{\rm c}$ and $F_{\rm
b}^{1/2}$ is acceptable except for Case I3. This may be due to the
more serious influence of radiation diffusion in I3 (the Peclet
number gets smaller). From the last two columns of Table~\ref{ovsh},
 one may infer a scaling ${v''}_{\rm zo}\propto F_{\rm b}^{1/3}$ which is not
consistent with ${v''}_z\propto F_{\rm b}^{0.25}$ given in
Section~\ref{sec:fluc}. In fact, ${v''}_z\propto F_{\rm b}^{1/3}$
works better than ${v''}_z\propto F_{\rm b}^{0.25}$ in the lower
overshooting zone. If we use the scaling relation $\Delta\propto
F_b^{1/2}$ to calculate the overshooting distance for $a_{\rm f}=1$,
the resulted value would be very small. However, the last column of
Table~\ref{ovsh} shows that the ratio $\Delta_{\rm c}/F_{\rm
b}^{1/2}$ increases as $F_{\rm b}$ decreases. Consequently, it is
possible that the overshooting distance is still substantial when
$F_b$ is close to be the realistic value. The penetration distance
for Case M1 is a direct simulation without scaling. It is comparable
to the value $0.63$ obtained by Xiong's convection model (XD2001).

Numerical simulations and Xiong's one-dimensional  stellar
convection theory generally give overshoot extents comparable to the
local $H_{\rm p}$. However, heiloseismological inference gives a
small overshoot extent, e.g., $\sim 0.1H_{\rm p}$(\cite{basu94}). As
pointed out in XD2001, the discrepancy may be caused by the
assumption of a break in the radial derivative of the sound speed.
The break is a consequence of the local MLT and does not appear in
Xiong's theory.

\subsection{Anisotropic Turbulence}

\begin{figure*}
  \centering
  \includegraphics[width=8.2cm]{./figures/xvz-ims10.eps}
  \includegraphics[width=8.2cm]{./figures/xvz-ms.eps}
  \caption{Common logarithms of $\chi$ (solid lines),
$\mathcal{Z}$ (dotted lines) and $|\mathcal{V}|$ (dashed lines) for
intermediate mass star model (Case I2, left panel) and massive star
model (Case M1, right panel). Double-headed arrow indicates the
overshooting zone.}
  \label{xvz}
\end{figure*}
\begin{figure*}
  \centering
  \includegraphics[width=8.2cm]{./figures/c3-ims.eps}
  \includegraphics[width=8.2cm]{./figures/c3-ms.eps}
  \caption{Height distributions of the anisotropic ratio
  ${w''}^2_z/({w''}^2_x+{w''}^2_y)$
of intermediate mass star models (left panel, solid line: Case I1;
dotted line: Case I2, dashed line: Case I3) and massive star model
(Case M1, right panel). Double-headed arrow indicates the
overshooting zone.}
 \label{c3s}
\end{figure*}

Xiong's (1977,1989a) non-local time dependent stellar convection
theory is a dynamical theory of auto- and cross-correlation
functions of the turbulent velocity and temperature fluctuations.
These fluctuations are defined as the derivations from the
density-weighted averages, namely,
\begin{equation}
u'_i=v_i-\frac{\overline{\rho v_i}}{\overline{\rho}},\quad
\tilde{T}'=T-\frac{\overline{\rho T}}{\overline{\rho}}.
\end{equation}
The starting point of Xiong's theory is a set of partial
differential equations for $\chi^2=\overline{w'_iw'^i}/3$,
$\mathcal{Z}=\overline{\tilde{T}'^2}/{\tilde{T}}^2$, and
$\mathcal{V}=\overline{\tilde{T}'w'_i}/{\tilde{T}}$, where
$w'_i=\rho u'_i/\overline{\rho}$, ${\tilde{T}}={\overline{\rho
T}}/{\overline{\rho}}$, and the summation convention for repeated
indices is used. The closure model contains three adjustable
parameters, $c_1$, $c_2$ and $c_3$ which describe the turbulent
dissipation, non-local turbulent diffusion, and anisotropy,
respectively. \cite{deng2006} stated that in the convectively
unstable region, the ratio of the vertical component to horizontal
component of motion is
${w''}^2_z/({w''}^2_x+{w''}^2_y)=(3+c_3)/2c_3$. In the upper
overshooting zone, ${w''}^2_z/({w''}^2_x+{w''}^2_y)\sim 0.5$ which
is independent of $c_3$. In the lower overshooting zone,
${w''}^2_z/({w''}^2_x+{w''}^2_y)\leq 0.5$ and it increases with
$c_3$.

The common logarithms of $\chi$, $\mathcal{Z}$ and $|\mathcal{V}|$
for the model cases are given in Fig.~\ref{xvz}. In both plots,
$\log{\mathcal{Z}}$ is the smallest one within the unstable zone,
and it becomes greater than $|\mathcal{V}|$ in the lower
overshooting layer. The downward dip of the dashed lines near the
bottom of the convective layer indicates the sign change of
$\mathcal{V}$ at the convective boundary. The distributions of
${w''}^2_z/({w''}^2_x+{w''}^2_y)$ for different models
 are given in Fig.~\ref{c3s}.
\cite{tian2009} (hereafter TDXC) and the current simulations suggest
that the spatial distribution of the anisotropic ratio
${w''}^2_z/({w''}^2_x+{w''}^2_y)$ is very sensitive to the
characters of the model. The maximum value generally occurs inside
the convection zone. From Fig.\ref{c3s} of the current study and
Fig.6 in TDXC, we can see that the maximum value of the anisotropic
ratio varies between $1.8$ to $2.8$. Case I1 $\sim$ I3 show that
larger input flux makes larger anisotropic ratio. For Case M1 in
which only a fraction of total energy flux is carried by convection,
the maximum anisotropic ratio is about $1.5$. Since
${w''}^2_z/({w''}^2_x+{w''}^2_y)=(3+c_3)/2c_3$ holds only in the
fully unstable convective region, we estimate $c_3$ inside the
convection zone. The values are approximately between $1$ and $3$.
In the lower overshooting layer of Case I3, the value of
${w''}^2_z/({w''}^2_x+{w''}^2_y)$ stays around $0.6$. This is not
compatible with the prediction of \cite{deng2006}. This could be
caused by the small size of the computed domain.

The anisotropy factor plays an important role in controlling the
total amount of acoustic energy injected into the solar oscillation
modes. For incompressible flows, the anisotropy factor adopted by
\cite{gough77} can be computed as ${w''}^2_z/({w''}^2_x+{w''}^2_y) =
1/(\Phi-1)$. In Gough's MLT, $\Phi=1.3745$ can match the observed
solar damping rates. On the other hand, B\"ohm-Vitense's MLT (1958)
requires a $\Phi=2$ which is close to the simulation results of
\cite{stein98}. Our results show that in the region where the
fluctuation of density is small, $\Phi$ is also very close to 2. So
it is possible to derive theoretical values for $\Phi$, such as
$\Phi=5/3$ for maximizing the convective heat flux (\cite{gough78}).

\subsection{Correlation Coefficients}

It is instructive to compare the numerical results with those
obtained by the 1D method of Xiong. Figure~\ref{cmprvt} compares the
correlation coefficients between the velocity and the temperature
fluctuations obtained by the different methods. Numerical
simulations give more extensive overshooting and lower values of
correlation in the convection zone. The larger overshooting may be
caused by the large kinetic energy flux which is negligible in the
Xiong's theory. The extended distributions of the kinetic flux need
to be balanced by similarly extended enthalpy fluxes. The
temperature-velocity correlation is proportional to the enthalpy
flux. Inside the convection zone, the numerical values of the
correlation coefficient are close to the one given in CS89 (0.81).

\begin{figure}
  \centering
  \includegraphics[width=8.2cm]{./figures/rvta-cmp.eps}
  \includegraphics[width=8.2cm]{./figures/rvtb-cmp.eps}
  \caption{Comparison of correlation coefficient
of temperature and velocity, i.e., $\mathcal{V}/(\chi_z
\mathcal{Z})^{0.5}$. (a) massive star model, (b) intermediate mass
star model. Solid lines: numerical results; dotted lines: data from
1D stellar model based on Xiong's convection theory. Note that in
the panel (b), the abscissas of solid line and dotted line are
scaled with different length ($0.1R$ and $0.37R$, respectively). In
the interior of realistic red giant,  the transition layer between
completely radiative zone and convection zone extends much deeper
than $0.1R$. The small humps at $z\approx 0.9$ correspond to the
upper convective boundaries. The wiggles near $z\approx 0.1$ in (a)
may be caused by the lower resolutions of numerical simulation.}
  \label{cmprvt}
\end{figure}

\subsection{Non-local Transport}
\label{nlt}

\begin{figure*}
  \centering
  \includegraphics[width=8.2cm]{./figures/nts-ims10.eps}
  \includegraphics[width=8.2cm]{./figures/nts-ms.eps}
  \caption{Some non-local turbulent transports for Case I2 and Case M1.
Left panel: Case I2, solid line: $NLT1\times 100$; dotted line:
$NLT2\times 1000$; dashed line: $NLT3\times 100$. Right panel: Case
M1, solid line: $NLT1\times 10$; dotted line: $NLT2\times 50000$;
dashed line: $NLT3\times 1000$. Double-headed arrow indicates the
overshooting zone.}
  \label{nts}
\end{figure*}

In Xiong's statistical turbulent convection theory, the non-local
transports are assumed to be of gradient type
(\cite{xiong89a,xiong98}):
\begin{eqnarray}
\label{ntl1}
NLT1&=&\overline{u'_k w'_i w'^i}=-\chi l_1 \nabla_k \overline{w'_i w'^i},\\
\label{ntl2}
NLT2&=&\overline{u'_k \tilde{T}'^2}/\tilde{T}^2=
-\chi l_3 \nabla_k (\overline{\tilde{T}'^2}/\tilde{T}^2),\\
\label{ntl3}
NLT3&=&\overline{u'_k w'^i \tilde{T}'}/\tilde{T}=-\chi
l_5 \nabla_k (\overline{w'^i \tilde{T}'}/\tilde{T}),
\end{eqnarray}
with $ l_1\simeq l_3 \simeq l_5=\Lambda$, where $\Lambda$ is the
Lagrangian integral length scale of turbulence. The parameters
introduced here, i.e., $l_1$, $l_3$ and $l_5$ are linked with $c_2$
and can be made dimensionless with $H_{\rm p}$. The numerically
obtained values of these non-local transports in our cases I2 and M1
are given in Fig.~\ref{nts}.

Estimations of $l_1/H_{\rm p}$, $l_3/H_{\rm p}$ and $l_5/H_{\rm p}$
based on Eqs.~(\ref{ntl1})$\sim$(\ref{ntl3}) are shown in
Fig.~\ref{lsims} and Fig.~\ref{lsms}.
\begin{figure}
  \centering
  \includegraphics[width=8.2cm]{./figures/l1-ims.eps}
  \includegraphics[width=8.2cm]{./figures/l3-ims.eps}
  \includegraphics[width=8.2cm]{./figures/l5-ims.eps}
  \caption{Non-local transport parameters for intermediate
mass star models. Pluses: Case I1; diamonds: Case I2; asterisks:
Case I3. Double-headed arrows indicate the overshooting zone. The
discontinuities in these height distributions are caused by either
singularity or numerical errors. The values near these
discontinuities are unreliable.}
 \label{lsims}
\end{figure}
\begin{figure}
  \centering
  \includegraphics[width=8.2cm]{./figures/l1-ms.eps}
  \includegraphics[width=8.2cm]{./figures/l3-ms.eps}
  \includegraphics[width=8.2cm]{./figures/l5-ms.eps}
  \caption{Non-local transport parameters for massive
star model. Double-headed arrow indicates the overshooting zone. The
discontinuities are induced by the reasons mentioned in the caption
of Fig.~\ref{lsims}.}
\label{lsms}
\end{figure}
There are no universal values for these parameters. The worst case
is the gradient of turbulent kinetic energy. $l_1/H_{\rm p}$ is far
from a constant. $l_3/H_{\rm p}$ is better and has a value around
$1.5$ in the unstable zone. It turns negative in overshooting layer.
$l_5/H_{\rm p}$ is roughly $5$ in the unstable zone and on the order
of $10$ in the overshooting zone. These parameters seem independent
of the details of the numerical models since they have nearly the
same profiles and magnitudes across the cases I1, I2 and I3. In the
rapidly varying region, the gradient approximations are too
imprecise to be acceptable.

\section{Summary and Conclusions}

In this paper, we have presented three-dimensional numerical
simulations of downward overshooting in the envelopes of massive and
intermediate mass giants. For the massive giant case, a 1D stellar
model of 15 $M\sun$ star was used as initial model. While for the
intermediate mass giant case, a 3 $M\sun$ star was mimicked
qualitatively. In our simulations, we adopted an artificially
modified OPAL opacity and treated radiative energy transport by the
diffusion approximation. The gas was regarded as fully ionized and
the radiation was included in the EOS.  In order to reduce the
thermodynamic relaxation timescale, the input energy fluxes of the
intermediate mass star models were enlarged by enhancing the
radiative conductivity. A parametric investigation of such
enhancement has been conducted. By statistical analysis of the
thermodynamically relaxed state, the properties of overshooting
below the convection zone were parameterized and compared with the
1D stellar model based on Xiong's non-local time-dependent turbulent
convection theory. The main results are summarized as follows.
\begin{enumerate}
  \item The relative fluctuations of thermodynamical variables such as density,
pressure and temperature are scaled by $F_{\rm b}^{0.5}$, e.g.,
$T''/\overline{T}\propto F_{\rm b}^{0.5}$,  and the velocity
fluctuations are scaled by $F_{\rm b}^{0.25}$. These are different
from the results based on simulations of efficient convection which
gave $T''/\overline{T}\propto F_{\rm b}^{2/3}$ and $v''_z\propto
F_{\rm b}^{1/3}$. The difference should be caused by the significant
presence of radiation energy transfer. Since the amplification
factor $a_{\rm f}$ is very large ($\sim 10^4$), the scaling
relationships here obtained are preliminary and need to be
investigated further.
  \item Even though the convective regions in our models are
quite shallow, the overshoot region are quite substantial ($1\sim 2$
PSHs). In the lower overshooting zone the temperature gradient is
superradiative.
  \item The scaling relations among penetration distance, input flux, and
vertical velocity, e.g., $\Delta_c\propto F_{\rm b}^{1/2}$,
$\Delta_c\propto {v''}_{zo}^{3/2}$ are acceptable in fitting current
numerical results but may not be applicable to the actual modeling
of intermediate mass giants.
  \item In the unstable region, the anisotropy ratio varies approximately
between $1$ to $3$.
  \item The non-local turbulent transports are not well described by
gradient models. No universal constant scaling parameters exist in
both the unstable and overshooting zones.
\end{enumerate}

It should be emphasized that the resolution and aspect ratio of our
numerical experiments are limited by the computational resources.
For intermediate mass star models, the radial dimension (in PSHs) of
the unstable region cannot cover the whole convection zone. The
effects of enhanced resolution and greater depths need to be studied
in the future steps.

\section*{Acknowledgments}

We thank the Department of Astronomy at Peking University for
providing computer time on their SGI Altix 330 system on which the
initial models were tested. We are also pleased to acknowledge use
of computer time provided by X. P. Wu and the invaluable
contributions of D. R. Xiong to the current research. This wok was
partially supported by the Chinese National Natural Science
Foundation (CNNSF) through 10573022. KLC thanks Hong Kong RGC for
support (project no. 600306).

\bsp

\label{lastpage}


\begin{thebibliography}{99}
\bibitem[\protect\citeauthoryear{Baker}{1987}]{baker87} Baker N. H., 1987,
in Hillebrandt W., Meyer-Hofmeister E., Thomas H.-C., eds, Physical
Processes in Comets, Stars, and Active Galaxies. Springer, Berlin,
p.105
\bibitem[\protect\citeauthoryear{Basu, Antia \& Narasimha}{1994}]{basu94} Basu, S., Antia, H. M., Narasimha, D.
, 1994, MNRAS, 267, 209
\bibitem[\protect\citeauthoryear{Brummell, Clune \& Toomre}{2002}]{brummell2002} Brummell N.H., Clune T.L., Toomre J, 2002,
ApJ, 570, 825
\bibitem[\protect\citeauthoryear{B\"ohm-Vitense}{1958}]{vitense58} B\"ohm-Vitense E.,1958, Astrophysik, 46, 108
\bibitem[\protect\citeauthoryear{Canuto}{1993}]{canuto93} Canuto V.M.,1993,
ApJ, 416,331
\bibitem[\protect\citeauthoryear{Chan \& Wolff}{1982}]{chan82} Chan K.L., Wolff
C.L., 1982, J.Comput. Phys, 47, 109
\bibitem[\protect\citeauthoryear{Chan \& Sofia}{1986}]{chan86} Chan K.L., Sofia
S., 1986, ApJ, 307, 222
\bibitem[\protect\citeauthoryear{Chan \& Sofia}{1989}]{chan89} Chan K.L., Sofia
S., 1989, ApJ, 336, 1022 (CS89)
\bibitem[\protect\citeauthoryear{Chan \& Sofia}{1996}]{chan96} Chan K.L., Sofia
S., 1996, ApJ, 466, 372
\bibitem[\protect\citeauthoryear{Deng, Xiong \& Chan}{2006}]{deng2006} Deng L.C., Xiong D.R., Chan K.L., 2006, ApJ, 643, 426
\bibitem[\protect\citeauthoryear{Deng \& Xiong}{2008}]{deng2007} Deng L.C., Xiong D.R., 2008,
MNRAS, 386, 1979
\bibitem[\protect\citeauthoryear{Emden}{1907}]{emden07} Emden R., 1907,{\it Gaskugeln} (Leipzig, Teubner)
\bibitem[\protect\citeauthoryear{Gough}{1977}]{gough77} Gough D.O., 1977, ApJ,
214,196
\bibitem[\protect\citeauthoryear{Gough}{1978}]{gough78} Gough D.O.,
1978, in Belvedere G., Paterno L., eds, Proc. EPS Workshop on Solar
Rotation, Catania Univ. Press, Sicily, Publ. No 162, p. 337
\bibitem[\protect\citeauthoryear{Graham}{1975}]{graham75} Graham E., 1975,
J. Fluid Mech., 70, 689
\bibitem[\protect\citeauthoryear{Keller \& Friedmann}{1924}]{keller24} Keller L.V.,
Friedmann A., 1924, in C.B. Biezeno C.B., Burgers J.M., eds,
Proceedings of the First International Congress on Applied
Mechanics, (Technische Boekhandel en drukkerij, J. Waltman, Jr.,
Delft), p. 395.
\bibitem[\protect\citeauthoryear{Kim \& Chan}{1998}]{kim98} Kim Y.C., Chan K.L., 1998, ApJ, 496, L121
\bibitem[\protect\citeauthoryear{Malagoli, Cattaneo \& Brummell }{1990}]{malagoli90} Malagoli A., Cattaneo F., Brummell N.H.,1990, ApJ, 361,L33
\bibitem[\protect\citeauthoryear{Pal, Singh \& Chan}{2007}]{pal2007} Pal P.S., Singh H.P., Chan
K.L., Srivastava M.P., 2007, Astrophys Space Sci,307,399
\bibitem[\protect\citeauthoryear{Porter \& Woodward}{2000}]{porter2000} Porter D.H., Woodward P.R.,
 2000, ApJS, 127, 159
\bibitem[\protect\citeauthoryear{Richtmyer and Morton}{1968}]{richtmyer1968} Richtmyer R.D., Morton K.W.,
 1968, Difference Method for Initial Value Problems(New York: Interscience).
\bibitem[\protect\citeauthoryear{Rogers, Glatzmaier \& Jones}{2006}]{rogers2006} Rogers T.M., Glatzmaier G.A., Jones C.A.,
 2006, ApJ, 653, 765
\bibitem[\protect\citeauthoryear{Rogers, \& Iglesias}{1992}]{rogers92} Rogers F.J., Iglesias C.A., 1992,
ApJS, 79, 507
\bibitem[\protect\citeauthoryear{Rogers, Swenson \& Iglesias}{1996}]{rogers96} Rogers F.J., Swenson F.J., Iglesias C.A., 1996,
ApJ, 456, 902
\bibitem[\protect\citeauthoryear{Roxburgh \& Simmons}{1993}]{roxburgh93} Roxburgh I.W., Simmons
J., 1993, A\&A, 277, 93
\bibitem[\protect\citeauthoryear{Roxburgh}{1998}]{roxburgh98} Roxburgh I.W., 1998, in Chan K.L., Cheng K.S., Singh H.P.,
 eds, ASP Conf. Ser. Vol. 138, Proc. 1997 Pacific Rim Conf. on Stellar Astrophysics. Astron. Soc. Pac., San Francisco, p.411
\bibitem[\protect\citeauthoryear{Saikia et al.}{2000}]{saikia2000} Saikia E., Singh H.P., Chan K.L., Roxburgh I.W., Srivastava M.P.,  2000, ApJ, 529, 402
\bibitem[\protect\citeauthoryear{Schmitt et al.}{1984}]{schmitt84} Schmitt J.H., Rosner R., Bohn H.U.,  1984, ApJ, 282, 316
\bibitem[\protect\citeauthoryear{Singh, Roxburgh \& Chan  }{1994}]{singh94} Singh H.P., Roxburgh I.W.,Chan K.L., 1994,
A\&A, 281, L73
\bibitem[\protect\citeauthoryear{Singh, Roxburgh \& Chan  }{1995}]{singh95} Singh H.P., Roxburgh I.W.,Chan K.L., 1995,
A\&A, 295, 703
\bibitem[\protect\citeauthoryear{Singh, Roxburgh \& Chan  }{1998}]{singh98} Singh H.P., Roxburgh I.W.,Chan K.L., 1998,
A\&A, 340, 178
\bibitem[\protect\citeauthoryear{Spiegel}{1971}]{spiegel71} Spiegel E.A., 1971,
ARA\&A, 9, 323
\bibitem[\protect\citeauthoryear{Stein \& Nordlund}{1989}]{stein89} Stein R.F., Nordlund \AA., 1998, ApJ, 342, L95
\bibitem[\protect\citeauthoryear{Stein \& Nordlund}{1998}]{stein98} Stein R.F., Nordlund \AA., 1998,
ApJ, 499, 914
\bibitem[\protect\citeauthoryear{Tian et al.}{2009}]{tian2009} Tian C.L.,
Deng L.C., Chan K.L, Xiong D.R., 2009, RAA, 9, 102  (TDXC)
\bibitem[\protect\citeauthoryear{Xiong}{1977}]{xiong77} Xiong D.R., 1977,
Acta Astron. Sinica, 18, 86
\bibitem[\protect\citeauthoryear{Xiong}{1985a}]{xiong85a} Xiong D.R., 1985a,
Scientia Sinica, 28,764
\bibitem[\protect\citeauthoryear{Xiong}{1985b}]{xiong85} Xiong D.R., 1985b,
A\&A, 150, 133
\bibitem[\protect\citeauthoryear{Xiong}{1986}]{xiong86} Xiong D.R., 1986,
A\&A, 167, 239
\bibitem[\protect\citeauthoryear{Xiong}{1989a}]{xiong89a} Xiong D.R., 1989a,
A\&A, 209, 126
\bibitem[\protect\citeauthoryear{Xiong}{1989b}]{xiong89b} Xiong D.R., 1989b,
A\&A, 213, 176
\bibitem[\protect\citeauthoryear{Xiong, Cheng \& Deng}{1998}]{xiong98} Xiong D.R., Cheng Q.L., Deng L.C., 1998,
ApJ, 500, 449
\bibitem[\protect\citeauthoryear{Xiong \& Deng}{2001}]{xiong01} Xiong D.R., Deng L.C., 2001,
MNRAS, 327, 1137 (XD2001)
\bibitem[\protect\citeauthoryear{Zahn}{1991}]{zahn91} Zahn J.-P., 1991,
A\&A, 252, 179
\end{thebibliography}
\end{document}